\documentclass[prd]{revtex4}
\newcommand{\be}{\begin{equation}}
\newcommand{\ee}{\end{equation}}
\newcommand{\bea}{\begin{eqnarray}}
\newcommand{\eea}{\end{eqnarray}}
\newcommand{\lb}{\label}
\newcommand{\I}{\mbox{i}}
\newcommand{\D}{\mbox{d}}
\newcommand{\E}{\mbox{e}}
\begin{document}
\title{ENTROPY OF GRAVITONS PRODUCED
            IN THE EARLY UNIVERSE}
\author{Claus Kiefer}
\affiliation{
 Fakult\"at f\"ur Physik, Universit\"at Freiburg,
  Hermann-Herder-Stra\ss e 3, D-79104 Freiburg, Germany.}
\author{David Polarski}
\affiliation{
 Lab. de Math\'ematiques et Physique Th\'eorique,
 UPRES A 6083 CNRS,
  Universit\'e de Tours, Parc de Grandmont, F-37200 Tours, France.}
\affiliation{
 D\'epartement d'Astrophysique Relativiste et de Cosmologie,
 Observatoire de Paris-Meudon, F-92195 Meudon cedex, France.}
\author{Alexei A. Starobinsky}
\affiliation{ 
 Landau Institute for Theoretical Physics, 
 Kosygina St. 2, Moscow 117940, Russia.}
\date{\today}
\begin{abstract}
Gravitons produced from quantum vacuum fluctuations during an
inflationary stage in the early Universe have zero entropy as
far as they reflect the time evolution (squeezing) of a pure state, 
their large occupation number notwithstanding. A non-zero entropy of the 
gravitons (classical gravitational waves (GW) after decoherence) can be
obtained through coarse graining. The latter has to be physically
justified {\it and} should not contradict observational constraints.
 We propose 
two ways of coarse graining for which the fixed temporal phase
 of each Fourier
mode of the GW background still remains observable: one based on
quantum entanglement, and another one following from the presence of a
secondary GW background. The proposals are shown to be mutually
consistent. They lead to the result that the entropy of the primordial GW
background is significantly smaller than it was thought earlier. The
difference can be ascribed to the information about the regular
(inflationary) initial state of the Universe which is stored in this
background and which reveals itself, in particular, in the appearance
of primordial peaks (acoustic peaks in the case of scalar perturbations)
in the multipole spectra of the CMB temperature anisotropy and
 polarization.    
\end{abstract}
\maketitle

\section{Introduction}

What is the entropy of primordial fluctuations from which all 
inhomogeneities in our Universe are assumed to originate? This 
fundamental problem becomes especially intriguing in the framework of the 
inflationary paradigm. Indeed, in that case primordial fluctuations 
are generated from vacuum quantum fluctuations. In this work, we
shall concentrate on a primordial gravitational-wave (GW) background
in inflationary cosmological models, although a similar discussion
can be applied to scalar perturbations in such models as well. It is
well known that in this case the unitary evolution of the quantum state of
the fluctuations outside the Hubble radius leads to an effective 
quantum-to-classical transition of a very specific kind: the
fluctuations become indistinguishable from classical 
fluctuations with stochastic amplitude and fixed temporal phase once
the so-called decaying mode of the perturbations is neglected~\cite{PS1}.
 
The approximation of simply omitting the decaying mode
 and then introducing
the equivalent classical stochastic fluctuation is sufficient
for calculating correctly the amplitudes
 of observable inhomogeneities in the 
Universe. It does not require any explicit account of interaction
with an environment. In particular, inflationary predictions for the
amplitude, statistics and quasi-classical temporal behaviour
 of perturbations
are independent of this interaction (at least, for sufficiently large
scales). Note that the possibility to get a
 decoherence-independent prediction
for observable quantities (inhomogeneities)
 is a specific property of the
very mechanism of generation of inhomogeneities
 in the inflationary scenario
of the early Universe, and does not occur
 for general quantum systems, as has 
been shown in \cite{PS1} with the help of the Wigner function
 and in \cite{KP,KLPS} by a 
discussion of thought experiments involving slits.
 Also, it may be attributed 
to the fact \cite{KPS} that amplitudes of perturbations
 constitute an (almost ideal) pointer
basis in this case. Finally, this approximation
 fits well into the consistent-histories
framework: as the decaying mode becomes negligible, probabilities can
be consistently assigned to classical trajectories (``histories'')
of perturbation modes in phase space \cite{DP}.

However, this does not mean that this approximation
 is sufficient for the
calculation of all quantities. In the present paper
 we consider one important
(though not directly measurable) quantity for which it is definitely
insufficient and for which
 one has to consider actual mechanisms of 
environment-induced decoherence --
 the entropy of cosmological inhomogeneities.
The entropies of an initial quantum perturbation
 in a pure strongly 
squeezed state and of its (approximate) equivalent 
(a classical perturbation 
with a stochastic amplitude, but a fixed phase) are
in fact both zero. It is, however, 
clear that, as a result of
 an effective coarse graining produced by
environment-induced decoherence, a pure quantum state
 becomes mixed, while the fixed 
classical phase of the classical system
 acquires some small stochastic part, so a non-zero entropy
should arise in both quantum and effective
 classical descriptions of perturbations. 

The main question, then, is: what is the correct coarse graining?
Any coarse graining introduced by hand must be justified by some
concrete physical mechanism. Also, it must be in agreement with 
present and future observational constraints. Indeed, as was already 
emphasized in~\cite{LPS}, the coarse graining that one chooses for
scalar (adiabatic) perturbations can be crucial with respect to the 
appearance, or not, of periodic oscillations - acoustic peaks - in the 
CMB anisotropy multipoles $C_{\ell}$. The same refers to the primordial GW
background where the corresponding periodic oscillations (with a different
period) are more easily seen in the CMB polarization multipoles 
$C_{Pl}$~\cite{PS1,LPPS}
(in this case, it is more appropriate  to call them 
primordial peaks). 

In earlier papers~\cite{Pr} where the entropy of cosmological 
perturbations was considered, an averaging procedure was performed with 
respect to the squeezing angle $\varphi_k$ of each Fourier mode,
leading to the entropy $S_k = 2r_k$ per mode, where $r_k$ is the
squeezing parameter. Such a ``high'' entropy would, however, 
spoil the generic prediction of the inflationary scenario about
the dominance of the quasi-isotropic (or growing, in the case of scalar 
perturbations) mode at the moment of the last Hubble-radius crossing,
which represents the equivalent classical expression
 of the remaining coherence 
between the field-mode amplitude and its momentum.
 As was shown in \cite{KLPS}, the
latter coherence persists long after
 the loss of quantum correlation between
quasi-isotropic and decaying modes of perturbations
 (and may survive even
up to the present moment for sufficiently
 large perturbation wavelengths). 
So, in the case of cosmological perturbations
 generated during inflation, the 
quantum decoherence sufficient for the classical description
 of the perturbations occurs long before the time of
 complete relaxation (if this relaxation occurs 
at all).
 
It is just this property that leads
to the appearance of the above-mentioned peaks in the CMB fluctuation
spectra. The advent of high-precision cosmological observations in 
general, and accurate measurements of the CMB fluctuations in particular, 
has dramatic consequences on this problem, which seemed remote only a
few years ago: the existence of peaks is incompatible with 
{\em complete} randomization of the temporal phase of the cosmological
fluctuations and severely restricts the choice of possible coarse 
grainings. 

Since, as is well known, the equations for small scalar (adiabatic)
perturbations and GW (tensor) perturbations superimposed on a 
Friedmann-Robertson-Walker (FRW) background can be reduced to
the equation of a real minimally coupled scalar field $\phi({\bf r},t)$, 
massless in the case of GW, while there is a mass term and a sound velocity 
in the case of adiabatic perturbations, we shall further consider this
auxiliary field. The FRW background is assumed to be spatially flat:
\be
ds^2=dt^2-a^2(t)dl^2~\ ,
\ee
where $dl^2$ is the three-dimensional
 Euclidean interval ($c=\hbar=1$ is assumed
throughout the paper). The quantum field 
$\hat\phi$ can then be decomposed into Fourier harmonics
\be
\hat\phi({\bf k})= \left(\phi({\bf k},t)\hat a({\bf k}) +
\phi^{\ast}(-{\bf k},t)\hat a^{\dagger}(-{\bf k})\right)e^{i{\bf
kr}} = \hat\phi^{\dagger}(-{\bf k})\ ,
\label{Fourier}
\ee
with $\hat a({\bf k})$ and $\hat a^{\dagger}({\bf k})$ being the usual 
annihilation and creation operators in the Fock space, respectively.

Let us take a two-mode system ${\hat\phi({\bf k}),\hat\phi(-{\bf k})}$.
This system exhibits a semiclassical behaviour by itself soon after
the first Hubble-radius crossing during an inflationary stage due to the 
disappearance of the decaying mode (in the Heisenberg representation), 
or equivalently due to large squeezing (in the Schr\"odinger
representation). More precisely, once the wavelength of the mode 
$\lambda = a(t)/k,~k\equiv {|\bf k|}$ 
is much bigger than the Hubble radius $R_H=H(t)^{-1}, ~H\equiv
\dot a/a$, we can make $\phi({\bf k},t)=\phi(k,t)$ real by a
time-independent phase rotation. Then, after the omission of the
decaying mode, the quantum mode operator 
(\ref{Fourier}) becomes equivalent to the Gaussian stochastic
quantity~\cite{PS1}
\be
\phi({\bf k})=\phi^{\ast}(-{\bf k})=c({\bf k})\phi(k,t)e^{i{\bf kr}}~,
\label{clas}
\ee
where $c({\bf k})= c^{\ast}(-{\bf k})$ is a complex Gaussian stochastic
field with the following non-zero correlations:
\be
\langle c({\bf k})c^{\ast}({\bf k'})\rangle = \delta({\bf k}-{\bf k'})~,
~\langle c({\bf k})c({\bf k'})\rangle = \delta({\bf k}+{\bf k'})~.
\ee

The stochastic field (\ref{clas}) has still zero entropy since its
volume in phase space is equal to zero. However, 
interactions with other fields, even tiny ones, are unavoidable.
This is a first physical process which can lead to coarse graining.
It occurs on very short decoherence timescales, rendering the 
field-amplitude basis a classical ``pointer basis''~\cite{KPS,KP} --
the robust basis with respect to the environment
\cite{Zurek,dec}. Once $\lambda \gg R_H$, 
the dynamical influence of such environmental fields should be 
negligible since causal processes are prohibited. 
In particular, the quasi-isotropic mode is not affected.
 However, for
decoherence it is just sufficient to destroy the quantum correlation
between the decaying and quasi-isotropic parts of perturbations.
 Since the 
decaying mode is ``local'' (in the sense that
 it may be affected by local 
processes), decoherence may be easily achieved
 by coarse graining of quantum
entanglement between (the decaying mode of) perturbations
 and the environment.
As was pointed out above,
the relaxation timescale exceeds by far the 
decoherence timescale (the latter being of the order of $H^{-1}$ at
the first Hubble radius crossing \cite{KP}).
 Since the maximum entropy $2r_k$ is 
associated with the relaxation timescale, the actual entropy arising
from entanglement, although non-vanishing, should be much smaller.

In the following we shall consider two different
coarse grainings which are physically relevant.
First, a realistic situation is considered where interaction with 
environmental fields is efficient
 in the suppression of off-diagonal terms
in the density matrix, but does not produce
 any significant back reaction
(Sec. II). Since it appears that details
 of the interaction are not relevant,
we model its effect through the introduction of a
 phenomenological parameter
suppressing the off-diagonal terms.
 Second, the entropy growth due to the 
loss of information about the primordial GW background 
which occurs as a result of the appearance of a secondary
GW background after the second Hubble radius crossing
 is calculated in Sec. III.
This effect arises because we cannot distinguish
 one kind of gravitons from
another at the present epoch.
 So, in the first case a non-zero entropy arises because
the primordial GW background is directly affected
 by an environment, while in the
second case it is simply ``polluted''
 by a secondary GW background emitted by some
environment.
It will be shown that both coarse grainings yield
in general $S_k$ 
significantly smaller than $2r_k$. Sec.~IV contains conclusions
and discussion, including the relation of this problem to the
problems of entropy growth in open systems and in the Universe.

\section{Entropy due to coarse graining of quantum entanglement}

Let us consider a squeezed vacuum state which describes
the behaviour of primordial fluctuations in the absence of interactions
with the environment. It has the form
\be
\psi_{k0}=\left(\frac{2\Omega_R}{\pi}\right)^{1/2}
        \exp\left(-[\Omega_R+\I\Omega_I]|y_k|^2\right)~,  \lb{psik0}
\ee
where $y_k$ is a Fourier component of the rescaled quantum field
$y({\bf r},t)=a(t)\phi({\bf r},t)$.
Since modes with different wave vectors ${\bf k}$ decouple,
we shall sometimes skip the index ${\bf k}$ (or $k$) in the following.
The wave function (\ref{psik0}) can be written in terms
of the squeezing parameters $r$ and $\varphi$ in the form
\be
\psi_0=\left(\frac{2k}{\pi[\cosh2r+ \cos2\varphi\sinh2r]}\right)^{1/2}
 \exp\left(-k\frac{1-\E^{2\I\varphi}\tanh r}
                  {1+\E^{2\I\varphi}\tanh r}|y|^2\right)\ . \lb{psi0}
\ee 
Squeezing can equivalently be expressed in terms of
``particle creation'' with average particle number $N({\bf k})\approx
\E^{2r}/4$. We are interested in the regime after the first
 Hubble-radius crossing (occurring during an inflationary stage) when 
$N({\bf k})\gg 1$.

To calculate the entropy it is sufficient
to consider one of the two modes contained in the complex
amplitude $y({\bf k},t)= y^{\ast}(-{\bf k},t)$ 
(in the following, $y$ is a real function).
The entropy is additive for the various modes.
The density matrix corresponding to the pure state (\ref{psik0})
is given in the field-amplitude representation by
\be
\rho_0(y,y')=\sqrt{\frac{2\Omega_R}{\pi}}
 \exp\left(-\frac{\Omega_R}{2}(y-y')^2-\I\Omega_I(y-y')(y+y')
    -\frac{\Omega_R}{2}(y+y')^2\right)\ . \lb{rho0}
\ee

The coefficient in front of $(y-y')^2$ is a measure of quantum coherence
(size of non-diagonal elements), while the coefficient in front of
$(y+y')^2$ is related to $(\Delta y)^2$ and
measures the extension in configuration space of a
fictitious ensemble described by the density matrix 
(details can be found in Appendix~A2.3 of \cite{dec}).
Since for large squeezing $\Omega_R$ becomes very small,
both coherence and extension in phase space become large.
This coherence cannot be distinguished from a classical random process,
see for instance the thought experiment involving slits in \cite{KP,KLPS,DP}. 
Coupling to environmental degrees of freedom \cite{KPS,KP} will decrease 
the coherence length in $y$, i.e. increase the coefficient in front of
$(y-y')^2$. Although details may be very complicated, a very good
approximation is to multiply $\rho_0$ by a Gaussian factor which suppresses
off-diagonal elements but leaves diagonal elements (probabilities)
untouched. This corresponds to the situation where dynamical
back reaction is small (``ideal measurement'').
Then, instead of (\ref{rho0}), the following density matrix is obtained:
\be
\rho_{\xi}(y,y')=\rho_0(y,y')\exp\left(-\frac{\xi}{2}(y-y')^2\right)\ ,
\lb{rhoxi}
\ee
where $\xi\gg\Omega_R$. For large squeezing, this condition reads 
\be 
\frac{\xi\E^{2r}}{k}\gg 1\ . \lb{dec}
\ee
In the following, it is referred to as the {\em decoherence condition}.

To study correlations that might be present between
$y$ and (the Fourier transform of) the momentum $p$, it is preferable to 
calculate the Wigner function. This is found to be
\be
W_{\xi}(y,p) 
= \frac{1}{\pi}\exp\left(-2\frac{(p+\Omega_Iy)^2}{(\Omega_R+\xi)}
    -2\Omega_Ry^2\right)\ . \lb{wigner}
\ee
The variances $\Delta y$ and $\Delta (p-p_{cl})$ can be found from the 
extension of the corresponding contour ellipse which describes
the extension of the (apparent) ensemble. 
This ellipse becomes a circle in the vacuum case ($r=0$, $\xi=0$), if we use 
the axes ${p}/{k},~y$.
It will be convenient to
introduce the variables ${\tilde y}\equiv 2\sqrt{k}y,~{\tilde p}\equiv 
\frac{2}{\sqrt{k}}p$, and we 
denote the major half axis of the Wigner ellipse in the 
${\tilde y},~{\tilde p}$ plane by
$\alpha$ and the minor half axis by $\beta$.
In the absence of environment ($\xi=0$), one has 
$\alpha_0=\E^r$, $\beta_0=\E^{-r}$ \cite{PS1}. 
For $\xi\neq 0$ one generally finds rather complicated expressions,
from which simple expressions are recovered in the limit of large squeezing.
One finds for $\E^{2r}\to\infty$
\be
\alpha\approx\E^{r}\ , \quad \beta\approx\sqrt{\frac{\xi}{k}}
           \gg \beta_0 \ . \lb{alphabeta}
\ee
While the size of the major half axis remains the same,
the size of the minor half axis becomes much bigger than before due
to decoherence. 

To preserve a correlation between $y$ and $p$ one has to demand
that $\alpha$ remains much bigger than $\beta$, or:
\be
\frac{\xi}{k\E^{2r}}\ll1\ . \lb{corr}
\ee
We shall refer to (\ref{corr}) as the {\em correlation condition}.
The surface of the ellipse (divided by $\pi$) is $A=\alpha\beta=
\E^r\sqrt{\xi/k}$. Thus, one would expect the following relation
between entropy and volume in phase space:
\be
S\approx\ln A\approx(1/2)\ln(\E^{2r}\xi/k)~. \lb{entr}
\ee 
This is what we shall demonstrate in the following. 

The von Neumann entropy connected with the density matrix
(\ref{rhoxi}) is given by
\be
S=-\mbox{Tr}(\rho_{\xi}\ln\rho_{\xi})\ . \lb{vN}
\ee
It was calculated in \cite{JZ} for an arbitrary Gaussian density
matrix, see also Appendix~A2.3 in \cite{dec}.
The result is
\be
S=-\ln p_0-\frac{q}{p_0}\ln q\ ,
\ee
where
\be
p_0=\frac{2\sqrt{\Omega_R}}{\sqrt{\Omega_R+\xi}+\sqrt{\Omega_R}}\ ,
\quad q=\frac{\sqrt{\Omega_R+\xi}-\sqrt{\Omega_R}}
             {\sqrt{\Omega_R+\xi}+\sqrt{\Omega_R}}\ .
\ee
Using $\E^r\to\infty$ and $\xi\gg\Omega_R$, one finds
\be
S\approx 1-\ln2+\frac{1}{2}\ln\frac{\E^{2r}\xi}{k}=1+
\frac{1}{2}\ln \frac{N\xi}{k}~, \lb{entropy}
\ee
in accordance with the expectation $S\approx\ln A$.
Applying the decoherence condition (\ref{dec}), one finds
\be
S\gg 1-\ln2\approx 0.31\ , 
\ee
where $\gg$ holds here in a logarithmic sense (it directly holds
for the number of states $\E^S$). Note that this lower bound on the
entropy corresponds to the loss of less than one bit of information.
This is consistent with previously known results on decoherence in 
quantum mechanics. For example, in recent quantum-optical experiments~
\cite{Ha}, decoherence starts if, on average, one photon is
lost. Thus, it may be expected that a minimal entropy
$S_{min}\approx\ln2$ per mode would be sufficient to guarantee
decoherence in the present case, too.

Applying the correlation condition~(\ref{corr}), we get
\be
S\ll 2r ,\lb{Smax}
\ee
where again $\ll$ holds in a logarithmic sense. It is evident
that the entropy must be much smaller than the maximum value
$2r$ which is found by integrating over the squeezing angle \cite{Pr}
(and which would just mean $\alpha=\beta\approx\E^{r}$ for the Wigner 
ellipse).

Note that, though our initial formula $S=\ln A$ (proposed earlier in~
\cite{LPS}, too) is the same as the one introduced
by Rothman and Anninos~\cite
{RA}, our final result~(\ref{Smax}) is drastically different from
the conclusion of the recent paper by Rothman~\cite{R99} that this formula
for the entropy leads to results identical to those obtained in~\cite{Pr}.
The reason for this difference is evidently the fact that in~\cite{RA,R99},
the total phase-space volume of a state with a given energy was 
calculated. This does not properly account for the squeezed nature of 
the Wigner ellipse in our case. 

Although details about $\xi$ may be complicated, it is natural to
expect that the coherence length $\xi^{-1/2}$ is not smaller than 
the width of the ground state for $r=0$, at least during the
inflationary stage, so that the quantum state really remains squeezed
in some direction as compared to the ground vacuum state. This means 
that $\xi<k$ and consequently $S<r$. Therefore, pure decoherence
(without dynamical influence) can never totally smear out
the Wigner ellipse to achieve $S\approx S_{max}=2r$. For complete
randomization, one would thus have to invoke, for example, a thermal
bath at a sufficiently high temperature. Such a model was 
discussed in \cite{KMH}, and
their results seem to be consistent with our general treatment.
Conditions similar to (\ref{dec}) and (\ref{corr})
have been frequently discussed in quantum mechanics, see e.g.
Eqs.~(6.32)-(6.36) in \cite{GMH}.

\section{Entropy due to a secondary GW background}

Let us consider now a different coarse graining due to the presence of a 
secondary GW background which is generated by different matter sources in a
causal way after a given mode of the fluctuations has re-entered the
Hubble radius~\cite{JAM,Allen}. For primordial GW, as shown in~
\cite{KLPS}, the coherence with respect to the
squeezing angle is maintained for a considerable time even
after the second Hubble-radius crossing. On the other hand,
 the secondary GW
background is expected to have a uniformly distributed phase. 
It is natural, and rather general, to assume that a ${\bf k}$-mode
of this secondary background is described by a density matrix 
$\rho^s({\bf k})$ which is diagonal in the 
occupation number basis, $\rho^s=\sum_n w_n|n\rangle\langle n|$ (here
the argument ${\bf k}$ is omitted, and $n=0,1...$).
This means in particular complete randomization of the temporal phase, or 
the absence of a 
preferred direction in phase space. As gravitons of the primary
 (squeezed) 
background are {\em indistinguishable} from those belonging
 to the secondary 
background, we expect some information loss when both
 backgrounds are mixed. 
The mean occupation number 
\be
n({\bf k})=\sum_{n=0}^{\infty}nw_n~,
\ee
though expected to be
significantly lower than that of the primary background $N({\bf k})$,
may nevertheless be large, too. We shall show that, loosely speaking, 
$\rho^s$ corresponds to (\ref{rhoxi}) with 
\be
\xi\simeq  \frac{16}{\E^2}kn({\bf k})~.
\label{corresp}
\ee

Let us estimate the corresponding entropy.
We consider the typical volume $\Gamma$ in (half of) phase space 
occupied by the system. For 
the isolated primary background, this volume $\Gamma_0$ is the minimal one 
which corresponds to zero entropy. The secondary background occupies a much 
larger volume which corresponds to circles in phase space 
with the approximate radius $\sqrt{\langle y^s~y^{s \dagger} \rangle}$, 
where
\be
k~\langle y^s~y^{s \dagger} \rangle = n({\bf k})+{1\over 2}~.
\ee
Here the averaging process involves also time averaging in addition to 
quantum average. Therefore at any time, the width of $\Gamma$ in the 
squeezed direction of the primary background will be given by the 
radius of the secondary background, while the elongated direction of the 
primary background remains dominated by the primary background. 
However, since the vacuum part $1/2$ should not be counted
twice (it has been already counted and ``squeezed'' in the primordial
part of the mode), only newly created gravitons $n({\bf k})$ produce
an additional noise in the initially squeezed direction in phase
space. If $N({\bf k})n({\bf k})\gg 1$, then $\Gamma$, the volume in
phase space occupied by the whole system, can be estimated as follows:
\be
\Gamma^2 \simeq \frac{k}{4}~\alpha_0^2~n({\bf k}) = 
N({\bf k})n({\bf k}) \gg \Gamma_0^2=\frac{1}{16}\ ,
\ee     
where $\Gamma_0$ is the phase-space
 volume of the initial vacuum state. The 
corresponding entropy $S({\bf k})$ can be estimated as 
\be
S({\bf k}) = \ln \frac{\Gamma}{\Gamma_0}
= r_k + \ln 2 + \frac{1}{2} \ln n({\bf k})
= r_k + \ln 2\pi -\frac{1}{2} \ln \frac{\omega^4}{\epsilon_c} + 
\frac{1}{2} \ln \Omega^s\lb{S}\ ,
\label{S2}
\ee 
where we have used the physical quantity $\Omega^s(\omega)\equiv 
\frac{\omega}{\epsilon_c}\frac{\D\epsilon^s}{\D\omega}
=\omega^4n_{\omega}/\pi^2\epsilon_c$
($\omega=k/a_0$)
which describes the secondary stochastic background, while $\epsilon_c$ 
denotes the critical density. 
Comparing (\ref{S2}) with (\ref{entropy}), the factor
 $16/\E^2$ is found in (\ref{corresp}). 
It follows that
the condition $N({\bf k})n({\bf k})\gg 1$ used above corresponds
to the decoherence condition (\ref{dec}), while the condition that the
secondary background is much smaller than the primordial one, 
$N({\bf k})\gg n({\bf k})$, corresponds to the correlation condition
(\ref{corr}).   

We see from (\ref{S}) that $S({\bf k})<2r_k$ as long as 
$n({\bf k})\ll N({\bf k})$, and also
$S({\bf k})\gg S^s\approx \ln n({\bf k})$ (for $n({\bf k})\gg 1$).
On large cosmological scales, we have 
$N({\bf k})\sim 10^{100}$, $r_k\sim 115$, and we have approximately 
for most models $N({\bf k})\propto k^{-4}$ for $10^{-16}\ll 
\nu=\omega/2\pi \ll 10^{10}$Hz. We expect that very little entropy
is produced by this mechanism on large scales, hence 
$S({\bf k})\ll 2r_k$ for these scales.

\noindent Several possibilities can be distinguished now.
\par\noindent
1) $n({\bf k})\ll 1$. \\ 
This case, certainly plausible on very large cosmological scales, 
yields $S({\bf k})\ll r_k$.  
\par\noindent
2) $n({\bf k})\approx 1$. \\ 
Now we have 
\be
S({\bf k})\approx r_k = {1\over 2}S_{max}~. 
\label{av}
\ee
Note, however, that $n({\bf k})\ll N({\bf k})$ and we still have a 
significant squeezing in phase space. Therefore, the standing-wave 
behaviour of the primordial GW still has definite observational 
consequences. 
\par\noindent
3) $1\ll n({\bf k})\ll N({\bf k})$. \\ 
In that case we have $r_k<S({\bf k})<2r_k$.
However, like for the preceding case, the system remains highly squeezed. 
Therefore, with respect to observations, this case is very similar to the 
preceding ones. 
\par\noindent
4) $n({\bf k})\geq N({\bf k})$. \\
Now, the squeezing of the primordial 
background is not apparent anymore in observations. When 
$n({\bf k})\gg N({\bf k})$, the squeezing is completely ``washed out''.

So, we have here another concrete coarse graining of the 
primordial GW stochastic background. For all cases for which 
$1\ll n({\bf k})\ll N({\bf k})$, which is a reasonable assumption on large 
cosmological scales, the resulting entropy is thus significantly smaller
than $2r_k$. The crucial point is that a 
value $r_k < S({\bf k}) < 2r_k$ is still fundamentally different from 
$S({\bf k})=S_{max}=2r_k$. In the first case, the composite system of both 
backgrounds can still reflect the squeezing of
 the primordial GW background, 
while it does not in the second case. 
The peaks in the B-mode polarization of the CMB produced by the total GW
background would be absent in the second case. 

\section{Conclusions and discussion}

We have shown that both methods of coarse graining proposed
yield for the entropy per mode
the result $S({\bf k})\ll 2r_k$
(in the logarithmic sense), which differs strongly from the results
obtained earlier~\cite{Pr}. The reason is that our coarse grainings do
not destroy the classical correlation
(the almost deterministic temporal
phase of the GW and of the adiabatic perturbations)
 remaining after quantum
decoherence, which, as physical estimates
show~\cite{KLPS}, persists for a rather long time (very long in the case
of GW) after the perturbations have re-entered the Hubble radius. 
Thus, though
the maximal entropy per mode $S({\bf k})=2r_k$ might eventually be reached 
after complete relaxation, it is not reached at the 
recombination time for sufficiently long-wave scalar perturbations
and even at the present time for sufficiently long-wave GW. As was
mentioned above, this leads to observable effects: periodic peaks in 
the multipole power spectra of the $\Delta T/T$-anisotropy and polarization 
of the CMB.
Since the difference between entropies arising after two coarse 
grainings of the same system may be interpreted as the difference of 
amount
of information loss due to these coarse grainings, the difference
between previous results for the entropy of primordial GW background and
our result may be interpreted as the amount of information contained in
primordial peaks (acoustic peaks in the case of scalar perturbations).

Note that the energy density of the primordial GW background calculated
either with the quantum operators ({\ref{Fourier}) or by the use of 
equivalent classical stochastic fields (\ref{clas}) is one half
of the energy density of the GW background described by 
a density matrix which is isotropic in 
phase space, with the same average number of gravitons given by
$n({\bf k})$ (this, of course, was taken into account in all correct
calculations of the energy density of GW produced during inflation,
beginning with~\cite{St79}). Now for the entropy, 
$S=S_{max}/2$ is also a distinguished, ``average'' case. However, 
we see that large deviations from $S_{max}/2$ on both sides are
possible; the entropy per mode can be both much larger or much
smaller than $r_k$ (in the logarithmic sense).

Let us finally discuss the temporal growth of the entropy of cosmological
perturbations, which is at least partly responsible for the arrow of 
time in the Universe \cite{zeh}.
 After the perturbations re-enter the Hubble radius,
$N({\bf k})$ remains constant (neglecting very small graviton-graviton
scattering events and graviton absorption by matter), so the growth of
entropy is mainly due to the growth of $n({\bf k},t)$ due to 
irreversible emission of gravitons by matter. Due to the expansion
of the Universe, $n({\bf k},t)$ typically grows only as a power of 
time, so $\dot S({\bf k},t)\propto t^{-1}$. The same refers to the
earlier period for $\lambda\gg R_H$, but after the end of inflation. 
(In inflation, one has for long wavelengths $\dot{S}\approx
H\approx\ constant$ and therefore $S\propto t$.)
Then, there is no secondary
GW background and we may use (\ref{entropy}) only. Under any
reasonable assumptions about the
 time dependence of $\xi$, $S({\bf k},t)$
grows logarithmically, so $\dot S({\bf k},t)\propto t^{-1}$, too.
Comparing this behaviour with those considered in the interesting
discussion~\cite{CCZP} of the paper~\cite{ZP}, we see that the
case of a two-mode subsystem of cosmological perturbations is
in some sense intermediate between the two different cases 
discussed in~\cite{CCZP} (decoherence due to a chaotic behaviour of the 
subsystem itself or decoherence caused by the environment).
Though our subsystem is not chaotic but only classically unstable,
the law for $\dot S({\bf k},t)$ is the same as for chaotic dynamical
systems.

\section*{Acknowledgements}

A.S. was partially supported by the Russian Foundation for 
Basic Research, grant 99-02-16224, and by the German Science
Foundation (DFG) through grant 436 RUS 113/333/5. A part of this 
paper was
made during his stay at the Institute of Theoretical Physics, ETH, 
Z\"urich. He thanks Prof. Ch. Schmid for hospitality there. C.K. is
grateful to the Institute of Advanced Study Berlin,
and the Isaac Newton Institute, Cambridge, for their kind hospitality
while part of this work has been done. D.P. acknowledges
C.K. for kind hospitality at the University of Freiburg.


\begin{thebibliography}{99}
\bibitem{PS1} D. Polarski and A.A.~Starobinsky,
 Class. Quantum Grav. {\bf 13}, 377 (1996).
\bibitem{KP} C. Kiefer and D. Polarski, Ann. Phys. (Leipzig) {\bf 7},
             137 (1998).
\bibitem{KLPS} C. Kiefer, J. Lesgourgues, D. Polarski, and
             A.A.~Starobinsky, Class. Quantum Grav. {\bf 15}, L67
             (1998).
\bibitem{KPS} C. Kiefer, D. Polarski, and A.A. Starobinsky,
              Int. J. Mod. Phys.~D {\bf 7}, 455 (1998).
\bibitem{DP} D. Polarski, Phys. Lett. B {\bf 446}, 53 (1999).
\bibitem{LPS} J. Lesgourgues, D. Polarski, and A.A.~Starobinsky,
              Class. Quantum Grav. {\bf 14}, 881 (1997).
\bibitem{LPPS} J. Lesgourgues, D. Polarski, S. Prunet, and
A.A.~Starobinsky, A\&A, submitted; preprint gr-qc/9906098.
\bibitem{Pr} R. Brandenberger, V. Mukhanov, and T. Prokopec, Phys. Rev. 
 Lett. {\bf 69}, 3606 (1992);
 T. Prokopec, Class. Quantum Grav. {\bf 10}, 2295 (1993);
 M. Gasperini and M. Giovannini, Phys. Lett. B~{\bf 301}, 334 (1993); 
 Class. Quantum Grav. {\bf 10}, L133 (1993);
 M. Kruczenski, L.E. Oxman, and M. Zaldarriaga, Class. Quantum Grav. 
 {\bf 11}, 2377 (1994).
\bibitem{Zurek} H.D. Zeh, Found. Phys. {\bf 3}, 109 (1973);
           W.H. Zurek, Phys. Rev.~D {\bf 24}, 1516 (1981). 
 \bibitem{dec} D. Giulini, E. Joos, C. Kiefer, J. Kupsch,
 I.-O. Stamatescu, and H.D. Zeh, {\em Decoherence and the Appearance
 of a Classical World in Quantum Theory} (Springer, Berlin, 1996).
\bibitem{JZ} E. Joos and H.D. Zeh, Z. Phys. B {\bf 59}, 223 (1985).
\bibitem{Ha} M. Brune, E. Hagley, J. Dreyer, X. Ma\^{\i}tre, A. Maali,
     C. Wunderlich, J.M. Raimond and S. Haroche, Phys. Rev. Lett.
     {\bf 77}, 4887 (1996).
\bibitem{RA} T. Rothman and A. Anninos, Phys. Rev.~D {\bf 55}, 1948
             (1997).
\bibitem{R99} T. Rothman, preprint gr-qc/9906002.
\bibitem{KMH} D. Koks, A. Matacz, and B.L. Hu, Phys. Rev.~D {\bf 55},
              5917 (1997).
\bibitem{GMH} M. Gell-Mann and J.B. Hartle, Phys. Rev. D {\bf 47}, 3345 (1993).
\bibitem{JAM} S. Bonazzola and J.-A. Marck, 
              Annu. Rev. Nucl. Part. Sci. {\bf 45}, 655 (1994). 
\bibitem{Allen} B. Allen, in: {\em Proceedings of the Les Houches School
         on Astrophysical Sources of Gravitational Waves}, edited by
         J.-A. Marck and J.-P. Lasota (Cambridge University Press, Cambridge,
         1997).
\bibitem{St79} A.A. Starobinsky, JETP Lett. {\bf 30}, 682 (1979).
\bibitem{zeh} H.D. Zeh, {\em The physical basis of the
 direction of time} (third edition, Springer, Berlin, 1999).
\bibitem{CCZP} G. Casati and B.V. Chirikov, Phys. Rev. Lett. {\bf 75},
               350 (1995); W.H. Zurek and J.P. Paz, {\it ibid}, 351.
\bibitem{ZP} W.H. Zurek and J.P. Paz, Phys. Rev. Lett. {\bf 72}, 2508
             (1994). 
\end{thebibliography}
\end{document}